\documentclass[twocolumn,superscriptaddress,nofootinbib,floatfix,aps,prb]{revtex4-1}
\usepackage{graphicx}
\usepackage{amsmath}
\usepackage{amssymb}
\usepackage{bm}
\usepackage{hyperref}
\usepackage{mathtools}

\usepackage{color}

\definecolor{orange}{rgb}{1,0.5,0}
\definecolor{col1}{RGB}{153, 52, 121}

\newcommand{\cG}{{\cal G}}

\newcommand{\ra}{\rangle}

\begin{document}

\title{
Isolated zeros in the spectral function as signature of a quantum continuum
}

\author{Nikolay Gnezdilov}
\author{Alexander Krikun}
\author{Koenraad Schalm}
\affiliation{Instituut-Lorentz, $\Delta$ITP, Universiteit Leiden, P.O. Box 9506, 2300 RA Leiden, The Netherlands}
\author{Jan Zaanen}
\affiliation{Instituut-Lorentz, $\Delta$ITP, Universiteit Leiden, P.O. Box 9506, 2300 RA Leiden, The Netherlands}
\affiliation{Department of Physics, Stanford University, Stanford CA 94305, USA}

\begin{abstract}
We study the observable properties of quantum systems which involve a quantum continuum as a subpart. We show in a very general way that in any system, which consists of at least two isolated states coupled to a continuum, the spectral function of one of the states exhibits an isolated zero at the energy of the other state. 
Several examples of quantum systems exhibiting such isolated zeros are discussed. 
Although very general, this phenomenon can be particularly useful as an indirect detection tool for the continuum spectrum in the lab realizations of quantum critical behavior.

\end{abstract}
\maketitle
 
\section{Introduction}
The energy levels of a generic quantum system can be organized either in discrete or continuum spectra. The discrete spectrum is associated with the existence of stable bound states, corresponding to localized long-lived quasiparticles with well defined energy, while the continuum spectrum reflects either multiparticle states 
with finite phase volume -- e.g., particle-hole continuum, a thermodynamically large number of interacting degrees of freedom -- a thermal bath, or the absence of quasiparticles as such -- a quantum critical continuum. The qualitative difference between discrete states and continua is manifest in the spectral function: It either exhibits sharp peaks in the former case or a (peakless) smooth profile in the latter one. In more complicated quantum systems which have both discrete and continuum subparts, the spectral function can take a distinct shape.
The well-known example is the Fano resonance, \cite{fano1961effects,zaanen1986strong,miroshnichenko2010fano} which arises when 
one probes the continuum in the presence of the isolated states, whose interference leads to a characteristic line shape with a neighboring peak and zero.

The continued strong interest in quantum criticality and various non-Fermi liquid models inherently concerns the study of a quantum system with a  continuum rather than a discrete spectrum. A defining feature of the non-Fermi liquid is the absence of the stable quasiparticles. No discrete peaks are seen in the spectral function; rather it has the appearance of a power law. \cite{varma1989phenomenology} The same type of spectra are characteristic of quantum critical systems, \cite{sachdev2011quantum} which are described by conformal field theory and a lack of the quasiparticles. Conformal field theories also generically exhibit power laws, controlled by the anomalous dimensions of the operators. These non-Fermi liquid and quantum critical ideas appear to be very relevant to the unconventional states of strongly correlated quantum matter, most notoriously the strange metal phase observed in high temperature superconductors and heavy fermion systems. \cite{keimer2015quantum}
This experimental relevance in turn triggered the active theoretical effort in the last decades, aimed at building controlled models of non-Fermi liquids and/or quantum critical systems. Among the latest developments is the Sachdev-Ye-Kitaev model, \cite{sachdev1993gapless,kitaev2015simple} which has a power law spectral function due to the strong entanglement between the constituent fermions. A few proposals have arisen recently on the experimental realization of the SYK model, \cite{chen2018quantum,pikulin2017black,chew2017approximating} which makes the question about the observable properties of the quantum critical continuum especially important. 
This case of the $0+1$ dimensional ``quantum dot'' systems is special since the multiparticle phase space shrinks to zero and any observed continuum spectrum can not be of quasiparticle nature. Therefore the detection of continuum in a quantum dot signals the interesting physics, whether it is related to the particular SYK model or not.

In this paper we point out a very basic and therefore very general feature of quantum systems containing the continuum as a subpart.  
We show that the spectral function exhibits a distinct line shape characterized by an isolated zero arising when one probes a discrete subpart of the system that consists both of discrete states and a continuum.
It is the ``mirror'' of the Fano resonance. A similar effect has been pointed out by one of us in the form of interference effects between 
different decay channels of the core holes created in high energy photoemission processes.\cite{zaanen1986strong} This distinct spectral function zero is in principle observable in experiment and it can therefore be used as a signature of the presence of a continuum subpart 
in quantum systems, in particular in the laboratory realizations of quantum criticality. 
The important aspect is that this probe is \textit{indirect} -- it does not interact with the continuum system. This can be a significant advantage since the quantum critical systems are notoriously fragile and the direct measurement could easily destroy them.

We shall first discuss the generic mechanism of the phenomenon and then demonstrate how it works in several examples with continuum subsystems: (1) two single fermion quantum dots coupled to an SYK quantum dot, (2) a one-dimensional wire coupled to a chain of the SYK nodes, and (3) a holographic model of a local quantum critical system with a periodic lattice.

\section{Isolated zeros in the spectral function} 
Consider two fermions $\chi_A, \chi_B$  with discrete quasiparticle energies $\Omega_A$, $\Omega_B$, respectively, coupled to a fermion $\psi$ with a continuum spectrum characterized by a Green's function $\mathcal{G}(\omega)$.  The Euclidean action for the full system reads:
\begin{align}
S&{}=S_\chi+S_\psi+S_{int} \label{S}, \\ \nonumber
S_\chi&{}=\int d t \sum_{\sigma=A,B} \bar{\chi}_\sigma \left(\partial_t +\Omega_\sigma\right)  \chi_\sigma, \\ \nonumber
S_\psi&{}=-\int dt dt' \, \bar{\psi}(t) \mathcal{G}\left(t-t'\right)^{-1} \psi(t'), \\ \nonumber
S_{int}&{}=\int dt \sum_{\sigma=A,B} \left(\lambda_\sigma \,  \bar{\psi} \, \chi_{\sigma}+\lambda_\sigma^* \bar{\chi}_{\sigma} \psi\right).
\end{align}
As shown in Appendix \ref{app_kappa} our effect is present for any tunneling couplings $\lambda_\sigma$, however for brevity here we consider the case $\lambda_A = \lambda_B \equiv \lambda$.
Integrating out the fermion $\psi$ in the continuum gives the Green's function for the fermions $\chi_A$, $\chi_B$:
\begin{gather} 
\hspace{-0.5em}
G_{\sigma\sigma'}(\omega)^{-1}\! \!=\!
\begin{pmatrix}
\omega \! - \! \Omega_A \! -\! |\lambda|^2 \mathcal{G}(\omega) & -|\lambda|^2 \mathcal{G}(\omega)\\
-|\lambda|^2 \mathcal{G}(\omega) &\omega \!- \! \Omega_B \!- \! |\lambda|^2 \mathcal{G}(\omega)
\end{pmatrix} \label{Gm1} \!.
\end{gather}
We are interested in the spectral function of a single fermion $\chi_A$: $A_{A}(\omega)=-\frac{1}{\pi}\mathrm{Im} G^R_{AA}(\omega)$, where the retarded Green's function is obtained from the inverse of the matrix (\ref{Gm1}) after analytic continuation $G^R(\omega)=G\left(\omega \to \omega+\mathrm{i} \delta \, \text{sign}(\omega)\right)$ with  $\delta=0^+$. 
The result is  
\begin{gather}
\label{DOSab}
A_{A}(\omega) = -\frac{|\lambda|^2}{\pi} 
\frac{\mathrm{Im} \mathcal{G}^R(\omega)}{\left|D(\omega)\right|^2}
\left(\omega-\Omega_B\right)^2,\\
\nonumber
D(\omega) \!= \! \left(\omega -\Omega_A\right)\!\left(\omega -\Omega_B\right) \!-\!|\lambda|^2 \!\left(2\omega -\Omega_A -\Omega_B\right)\!\mathcal{G}^R(\omega).
\end{gather}
In \eqref{DOSab} $\mathcal{G}^R$ is the retarded Green's function of the fermion $\psi$.
Here we've taken the limit $\delta\rightarrow 0^+$ assuming the imaginary part of $\mathcal{G}^R$ stays finite. In Eqs. (\ref{appA_GR}) and (\ref{appB_GR}) in Appendices \ref{app_kappa} and \ref{app_SYK4} we give the results for finite $\delta$, i.e., a universal finite lifetime for the fermions $\chi_{A,B}$.

\textbf{Our main observation} is that 
the spectral function of fermion $\chi_A$ has a double zero exactly at the energy level of fermion $\chi_B$: $\omega=\Omega_B$. 
The physics of the zero in \eqref{DOSab} can be better understood by considering the matrix structure of the imaginary part of the Green's function 
\begin{align}
\hspace{-0.5em}
\mathrm{Im} G^R_{\sigma \sigma'} \! \sim \!
&\begin{pmatrix}
 \left(\omega-\Omega_B\right)^2 & \! \left(\omega-\Omega_B\right) \! \left(\omega-\Omega_A\right) \\
  \! \left(\omega-\Omega_B\right) \! \left(\omega-\Omega_A\right) & \left(\omega-\Omega_A\right)^2
  \end{pmatrix}. 
\end{align}
This matrix is degenerate and has a single zero eigenvalue with eigenvector
\begin{equation}
\label{equ:chi}
|\chi_0\ra = (\omega-\Omega_A)|\chi_A\ra  - (\omega - \Omega_B)|\chi_B\ra ~.
\end{equation}
In other words this linear combination of probes is completely oblivious to the continuum. One immediately sees that for $\omega=\Omega_B$ the probe fermion $\chi_A$ aligns with the ``oblivious'' combination and thus does not get absorbed/reflected. This explains the double zero at the resonance frequency. In the degenerate case $\Omega\equiv\Omega_A=\Omega_B$ the double zero at $\omega=\Omega$ is canceled by the same multiplying factor in the denominator $D(\omega)$ and it's easy to understand the absence of the effect: For any energy the combination \eqref{equ:chi} is now fixed to $|\chi_0\ra \sim |\chi_A\ra  - |\chi_B\ra$ and it never coincides with one of the probes.

Though similar, this occurrence of the double zero for the discrete probe is the obverse of the Fano resonance, as the latter follows when one integrates out the discrete fermions $\chi_{A,B}$. The Green's function for the continuum fermion $\psi$ then becomes (setting the $\chi_B$ coupling to zero)
$G_{{\psi}\psi}^{-1} = {\cal G}^{-1} - {|\lambda|^2}/{(\omega-\Omega_A)}$,
with the spectral function 
\begin{equation}
\hspace{-0.5em}
A_\psi(\omega) \! = \! -\frac{1}{\pi}\mathrm{Im}G^R_{\psi \psi}(\omega) \! = \! -\frac{1}{\pi} \frac{(\omega - \Omega_A)^2 \mathrm{Im} \mathcal{G}^R(\omega)}{|\omega - \Omega_A - \lambda^2 \mathcal{G}^R(\omega) |^2}.
\end{equation}
The double zero at the energy $\omega=\Omega_A$ is again obvious but the pole structure is different. This Fano response has a pole at  $\omega = \omega_p$ near $\Omega_A$ shifted by the interaction with continuum.
For small $\lambda$ we can approximate its location as $\omega_p = \Omega_A + \lambda^2 \mathcal{G}(\Omega_A)$. In this case the spectrum function near the pole $\omega \sim \omega_p$ takes the familiar Fano form \cite{miroshnichenko2010fano}
\begin{equation}
A_\psi \propto \frac{(\epsilon + q\Gamma)^2}{\epsilon^2 + \Gamma^2},
\end{equation}
with the parameters 
$\epsilon = (\omega - \Omega_A - \lambda^2 \text{Re} \cG(\Omega_A))$, \mbox{$\Gamma =\lambda^2 \mathrm{Im} \mathcal{G}(\Omega_A)$}, and $q=\text{Re}\cG(\Omega_A)/\text{Im}\mathcal{G}(\Omega_A)$. 

One obvious extension of the action is to include a direct coupling between fermions $S_\chi'=\int dt\left(\kappa\bar{\chi}_A \chi_B +\kappa^*\bar{\chi}_B \chi_A\right)$. We analyze this in Appendix \ref{app_kappa}. 
The complex zeros of the spectral function $A_{A}(\omega)$ are given in this case by $(\Omega_B - \mathrm{Re}\kappa \pm \mathrm{i} \mathrm{Im}\kappa)$. The real part of the coupling shifts the position of the zero, while the imaginary one moves it off the real axis. In the latter case the exact real zero in the spectral function gets superseded by the localized depression with finite minimum value, but the overall line shape stays intact. This is reminiscent of the Fano resonance, where the exact zero is never observed in practice, but the full line shape points out the characteristic physics. \cite{zaanen1986strong,miroshnichenko2010fano}

\section{SYK model}

\begin{figure}[b!]
\center
\begin{minipage}{0.6 \linewidth}
\includegraphics[width=1\linewidth]{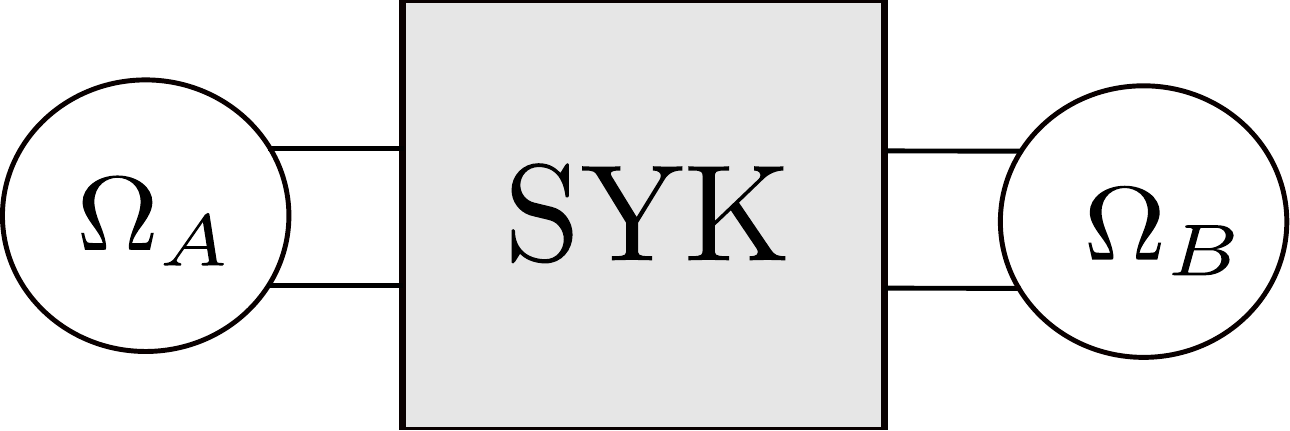}
\end{minipage}
\caption{\small \label{fig:chebu} \textbf{Cheburashkian geometry}. \cite{uspenskiy} Two quantum dots with descrete energy levels $\Omega_A$ and $\Omega_B$ are symmetrically coupled to an SYK quantum dot.}
\end{figure}

\begin{figure*}[ht!]
\center
\includegraphics[width=1 \linewidth]{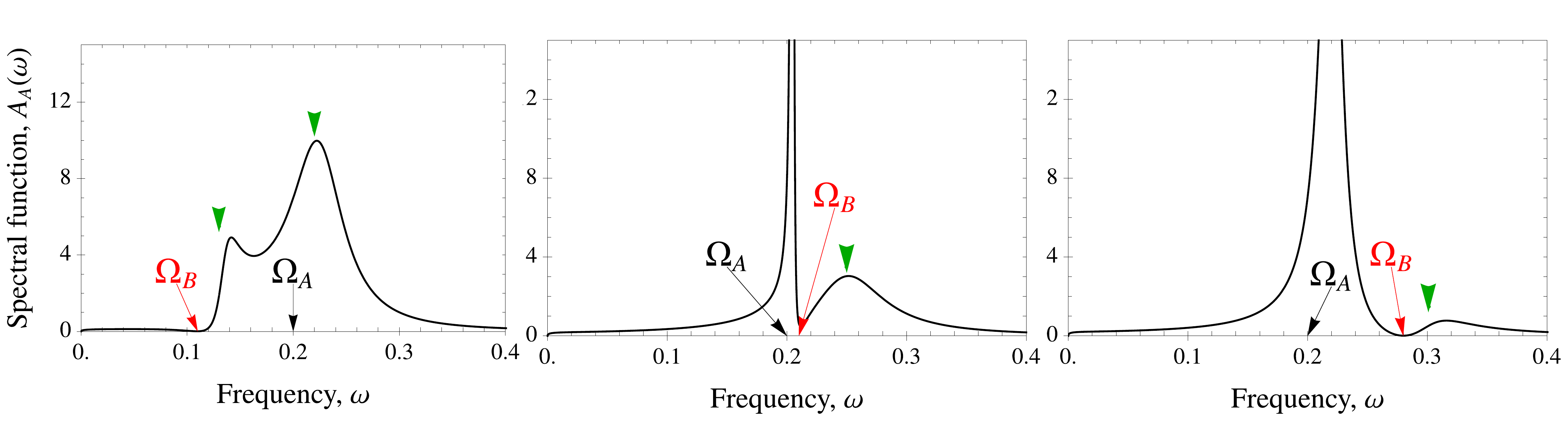}
\caption{\small \label{fig:position_scan} \textbf{Isolated zero for the SYK model}. For a given coupling to a continuum $\lambda=0.135, F=10$, the dependence of the spectral function of state $A$ on the energy level of the state $B$ is shown. The isolated zero is always present at $\Omega_B$. The peak at $\Omega_A$ gets sharpened due to the proximity of zero, but is destroyed if $\Omega_A=\Omega_B$ [not shown, see Eq. \eqref{DOSab}]. Green arrows show the positions of the poles of the Green's function defined as zeros of the determinant $D(\omega)$ in the spectral function in Eq. \eqref{DOSab}. Note that they do not coincide exactly with the maxima of the spectral function due to the proximity of zeros.}
\end{figure*}

\begin{figure*}[ht!]
\center
\includegraphics[width=1 \linewidth]{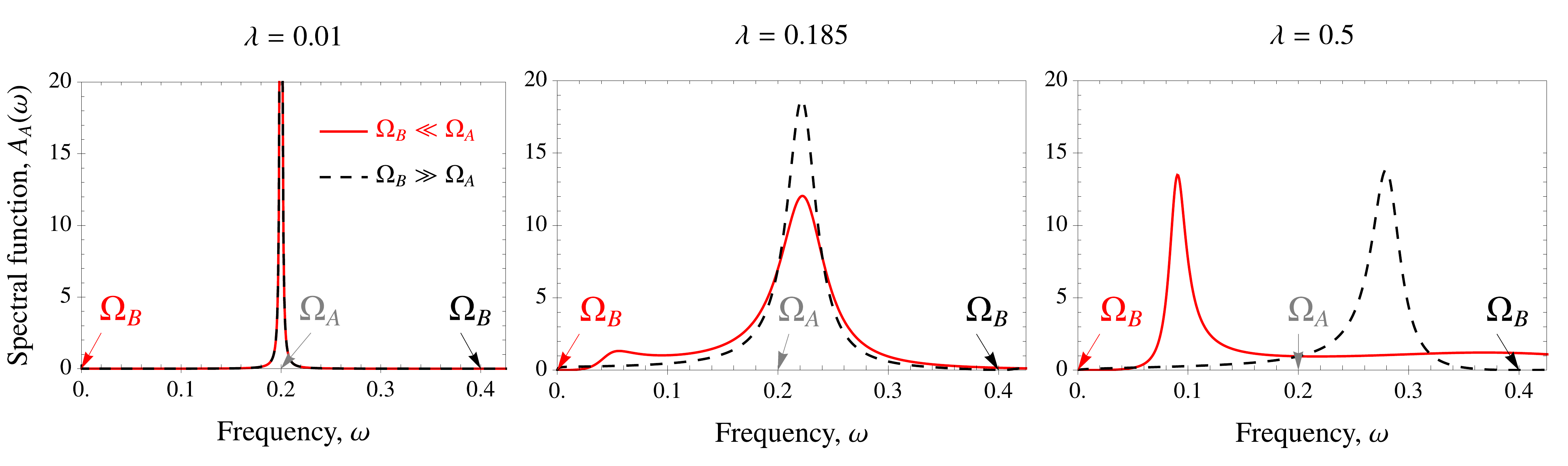}
\caption{\small \label{fig:symptote} \textbf{Asymptotic shape of the peak}. Depending on the value of the coupling $\lambda$ the shape of the peak at $\Omega_A$ may differ considerably in the asymptotic cases when $\Omega_B \ll \Omega_A$ (red lines) and $\Omega_B \gg \Omega_A$ (black lines). When $\lambda \ll 1$ (left panel) the two cases are identical, for intermediate values of $\lambda$ (middle panel) the width of the peak changes, for strong coupling (right panel) the position of the peak is affected as well. 
}
\end{figure*}

To give an \textbf{explicit example} of how this zero arises we first focus on the $0+1$ dimensional model featuring quantum critical continuum --
the Sachdev-Ye-Kitaev (SYK) model \cite{sachdev1993gapless,kitaev2015simple} with quartic interaction among complex fermions. \cite{sachdev2015bekenstein} We couple the continuum SYK model at charge neutrality point ($\mu_{SYK}=0$) to two discrete states with combined Hamiltonian
\begin{align}
H&{}=\sum_{\sigma=A,B} \Omega_\sigma \chi^\dag_\sigma \chi_\sigma+H_{SYK}+H_{int} \label{H}, \\ \nonumber
H_{SYK}&{}=\frac{1}{(2 N)^{3/2}} \sum_{i,j,k,l=1}^N J_{ij;kl} \,  \psi^\dag_i \psi^\dag_j \psi_k \psi_l, \\ \nonumber
H_{int}&{}=\frac{1}{\sqrt{N}}\sum_{i=1}^N \sum_{\sigma=A,B} \left(\lambda_i \psi^\dag_i \chi_\sigma+\lambda_i^* \chi^\dag_\sigma \psi_i\right).
\end{align} 
The fermions $\chi_A$, $\chi_B$ can be thought of as a pair of single state tunable quantum dots symmetrically coupled to the SYK system, the latter might be experimentally realized in a graphene flake based device. \cite{chen2018quantum}
We call it Cheburashkian geometry, \cite{uspenskiy} Fig. \ref{fig:chebu}.
The random couplings $J_{ik;kl}$ and $\lambda_i$ are independently distributed as Gaussian with zero mean $\overline{J_{ij;kl}}=0=\overline{\lambda_i}$ and finite variance $\overline{\left|J_{ij;kl}\right|^2}=J^2$ 
($J_{ij;kl}=-J_{ji;kl}=-J_{ij;lk}=\left(J_{kl;ij}\right)^*$), $\overline{|\lambda_i|^2}=\lambda^2$.
After disorder averaging, one finds that in the large $N$, long time limit $1 \ll J \tau \ll N$ the spectral function $A_A(\omega)$ for the fermion $\chi_A$ coincides with Eq. (\ref{DOSab}), where the coupling strength is given by the variance $\overline{|\lambda_i|^2}=\lambda^2$ and the continuum $\mathcal{G}^R$ originates from the SYK Green's function
\begin{align}
\mathcal{G}^R(\omega)=-\mathrm{i} \pi^{1/4}\frac{\mathrm{e}^{\mathrm{i} \pi \mathrm{sgn}(\omega)/4}}{\sqrt{J|\omega|}} \label{GR_SYK}
\end{align}
at zero temperature, see the details in Appendix \ref{app_SYK4}.

The line shape of the spectral function of $\chi_A$ depends on the frequencies $\Omega_A$ and $\Omega_B$ in a characteristic manner. We give examples in Figs. \ref{fig:position_scan} and \ref{fig:symptote}.
By tuning those one gets a distinct identifier of existence of the SYK continuum spectrum.

In case of multiple $M$ states, a spectral function of a single state $A_A(\omega) \propto \mathrm{Im} \mathcal{G}^R(\omega)\, \prod_{m=1}^{M-1} \left(\omega-\Omega_{B_m}\right)^2$ contains $M-1$ isolated zeros. This is described in Appendix \ref{app_ManyStates}, where we assume $M\ll N$ to avoid a transition in the SYK model to a Fermi liquid phase. \cite{banerjee2017solvable}

\section{Cluster of SYK nodes}

\begin{figure}[b!]
\center
\begin{minipage}{0.9 \linewidth}
\includegraphics[width=1\linewidth]{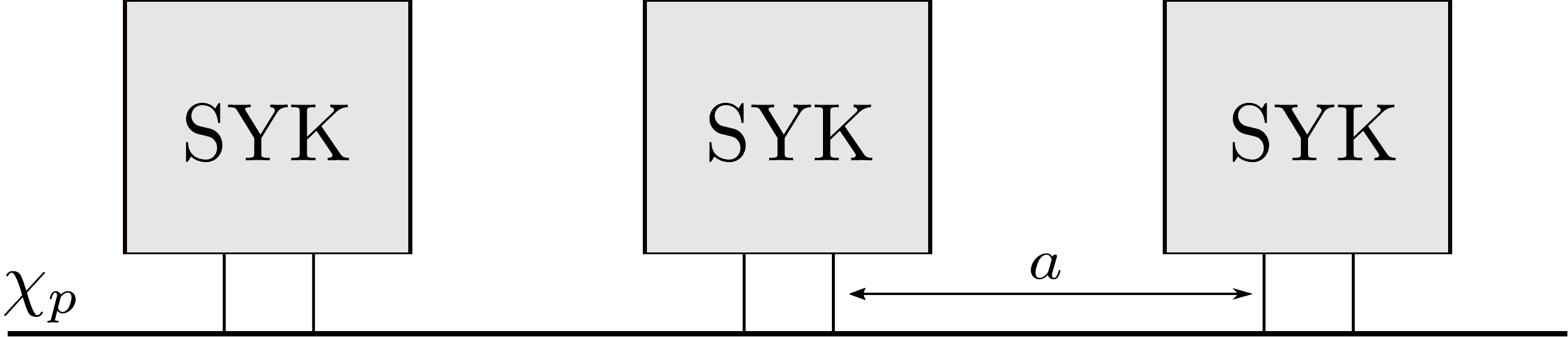}
\end{minipage}
\caption{\small \label{fig:chain} \textbf{SYK chain}. An evenly spaced chain of SYK impurities is introduced to the one-dimensional system with propagating fermion $\chi$. }
\end{figure}

One can generalize the previous case by clustering evenly separated quantum SYK dots, coupled to a 1D wire. \cite{ben2018strange,patel2018magnetotransport}
Here we can consider the itinerant fermion $\chi_p$  either in Galilean continuum or in an independent crystal with arbitrary periodicity. It has a given dispersion relation $\xi_p$ and interacts locally with an SYK quantum dot at point $x_n$, see Fig. \ref{fig:chain}.
The corresponding Hamiltonian is similar to Eq. (\ref{H}) with
\begin{gather}
H{}=\sum_{p} \xi_p \chi^\dag_p \chi_p+\sum_{x_n} H_{SYK}(x_n)+H_{int}(x_n) \label{H_chain}, \\ \nonumber
\begin{split}
H_{SYK}(x_n){}& \! =\! \frac{1}{(2 N)^{3/2}} \sum_{i,j,k,l=1}^N J_{ij;kl}^n \,  \psi^\dag_{i,n} \psi^\dag_{j,n} \psi_{k,n} \psi_{l,n}, \\ \nonumber
H_{int}(x_n){}& \!=\! \frac{1}{\sqrt{N}} \! \sum_{i=1}^N \! \left(\lambda_{i,n} \psi^\dag_{i,n} \chi(x_n)+\lambda_{i,n}^* \chi^\dag(x_n) \psi_{i,n} \right)\!.
\end{split}
\end{gather} 
It is clear that momentum plays now the role of the quantum number $A,B$ in our earlier model \eqref{S}. Therefore we are dealing with the case of many coupled states $M>2$. Importantly however, the momentum modes are not all cross coupled through the interaction mediated by many SYK continua. As usual for the periodic structures, see Appendix \ref{app:umklapps}, the SYK chain introduces an Umklapp-like effect: After integrating out all the SYK fermions the effective interaction only couples the momenta separated by the integer number of the reciprocal lattice units $\Delta p = 2\pi/a$. Note that only a discrete set of momentum modes is coupled, therefore the SYK fermions stay in the quantum critical phase.
The inverse Green's function is labeled by the Bloch momentum $p$
\begin{align} \nonumber
&{}G(p, \omega)^{-1}=\\
&{}=
\label{Gpp}
\begin{pmatrix}
\ddots & \cdots & \cdots \\
\vdots & \omega \! -\xi_p \! -\lambda^2 \mathcal{G}(\omega) & -\lambda^2 \mathcal{G}(\omega) \\
\vdots & -\lambda^2 \mathcal{G}(\omega) &\omega \! -\xi_{p - \Delta p} \! -\lambda^2 \mathcal{G}(\omega) 
\end{pmatrix}.
\end{align}
Following the same logic as before we find that the energy distribution of the spectral density at given momentum $p$ has a discrete set of zeros at frequencies $\omega = \dots, \xi_{p-\Delta p}, \xi_{p+\Delta p}, \xi_{p+ 2\Delta p}, \dots$. Varying $p$ this introduces continuous lines of zeros separated by $\Delta p$ in the momentum resolved spectral function $A(p,\omega)$, as seen in Fig. \ref{fig:ChainFig}, left panel. 
One can consider disordering the periodic structure by randomly shifting the nodes. The sharp lines in the Brillouin zone will smear out, since the Bloch momentum will not be well defined anymore. Therefore we expect the lines of zeros to smear out as well, but for the weak disorder the characteristic line shape will remain visible.

\section{Holographic fermions} 

\begin{figure}[b!]
\center
\includegraphics[width=0.6 \linewidth]{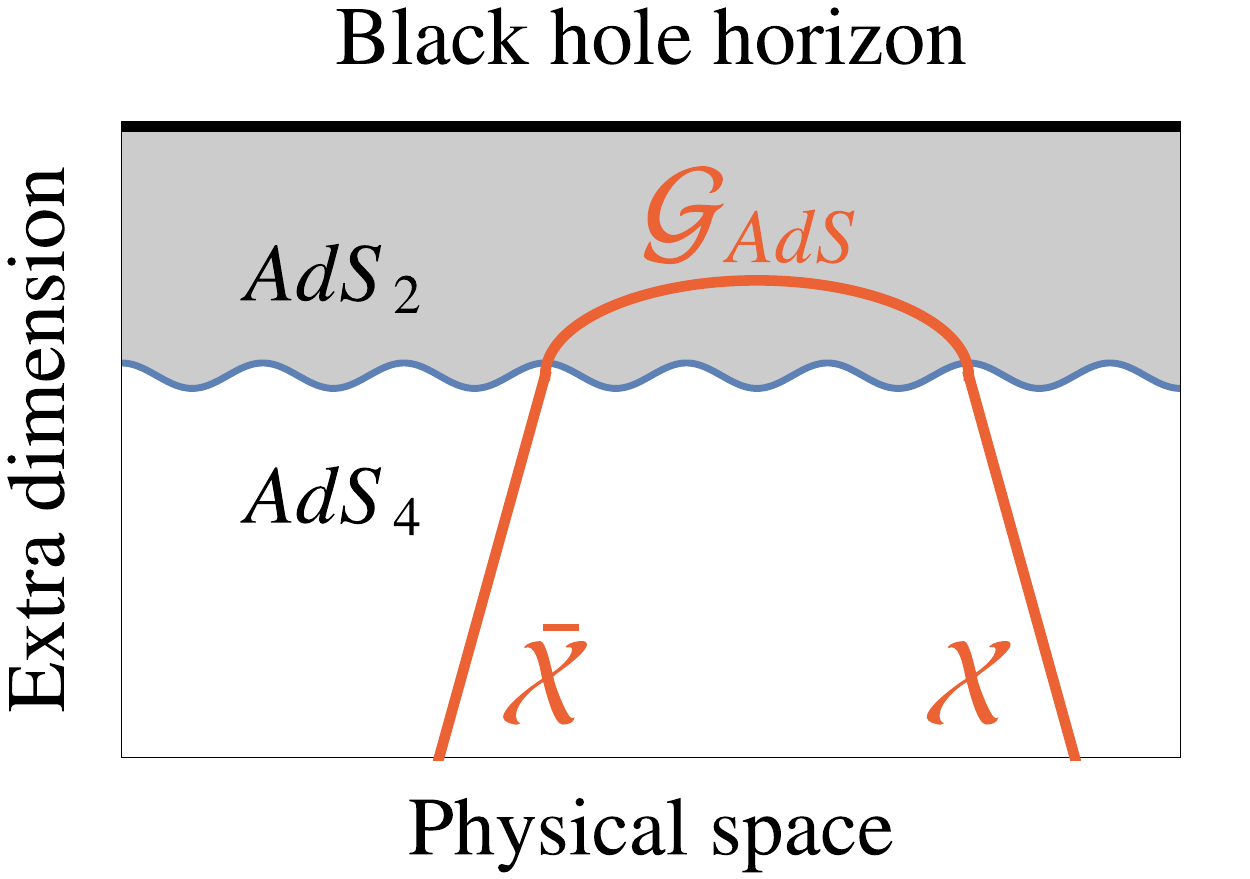}
\caption{\small \label{fig:holo} \textbf{Holographic fermions}. The spectral function of the fermion corresponds to the propagator in the holographic model. It is affected by the near horizon geometry, which induces locally critical continuum contribution $\mathcal{G}_{AdS}$.}
\end{figure}

\begin{figure*}[t!]
\center
\includegraphics[width=0.48 \linewidth]{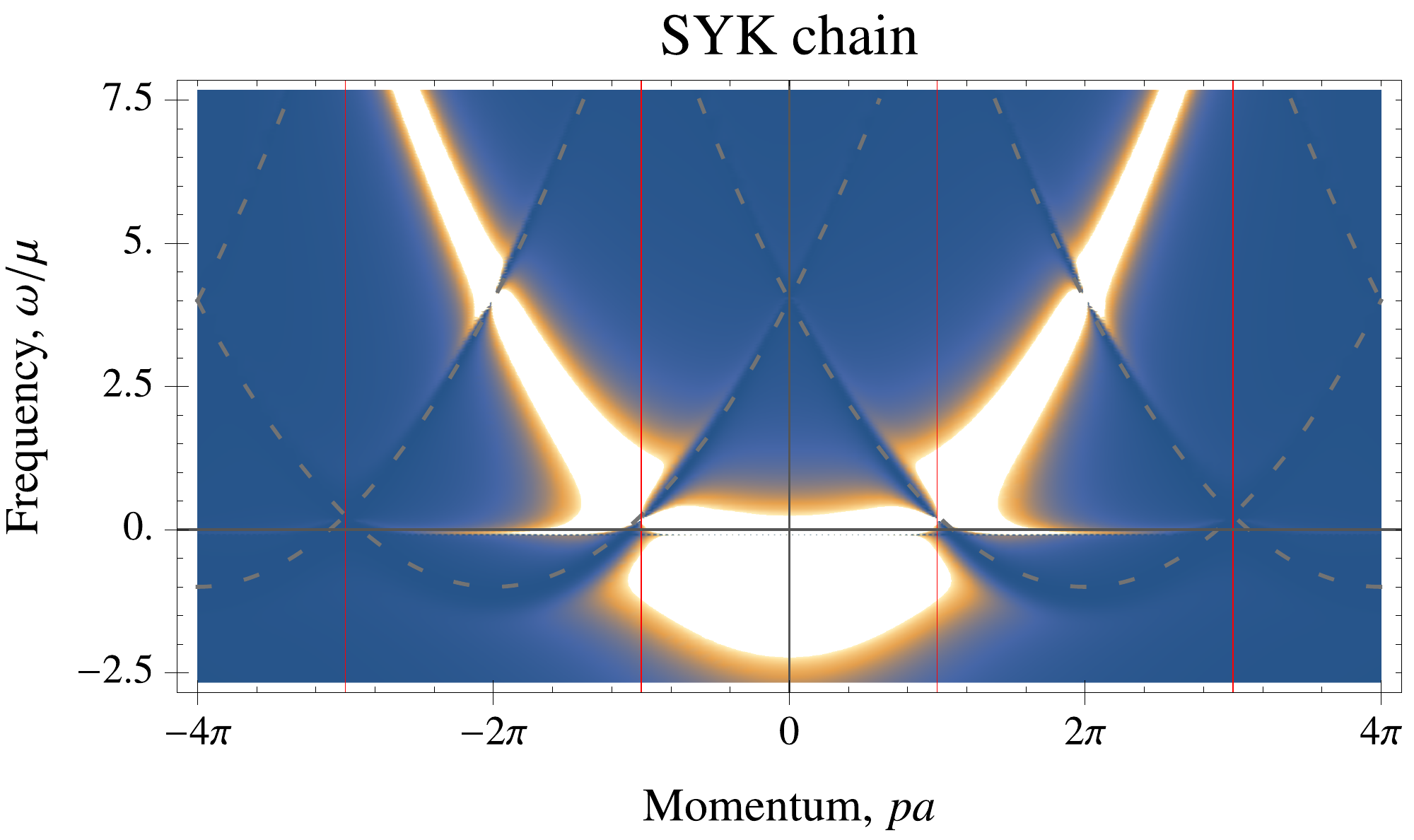} \quad
\includegraphics[width=0.48 \linewidth]{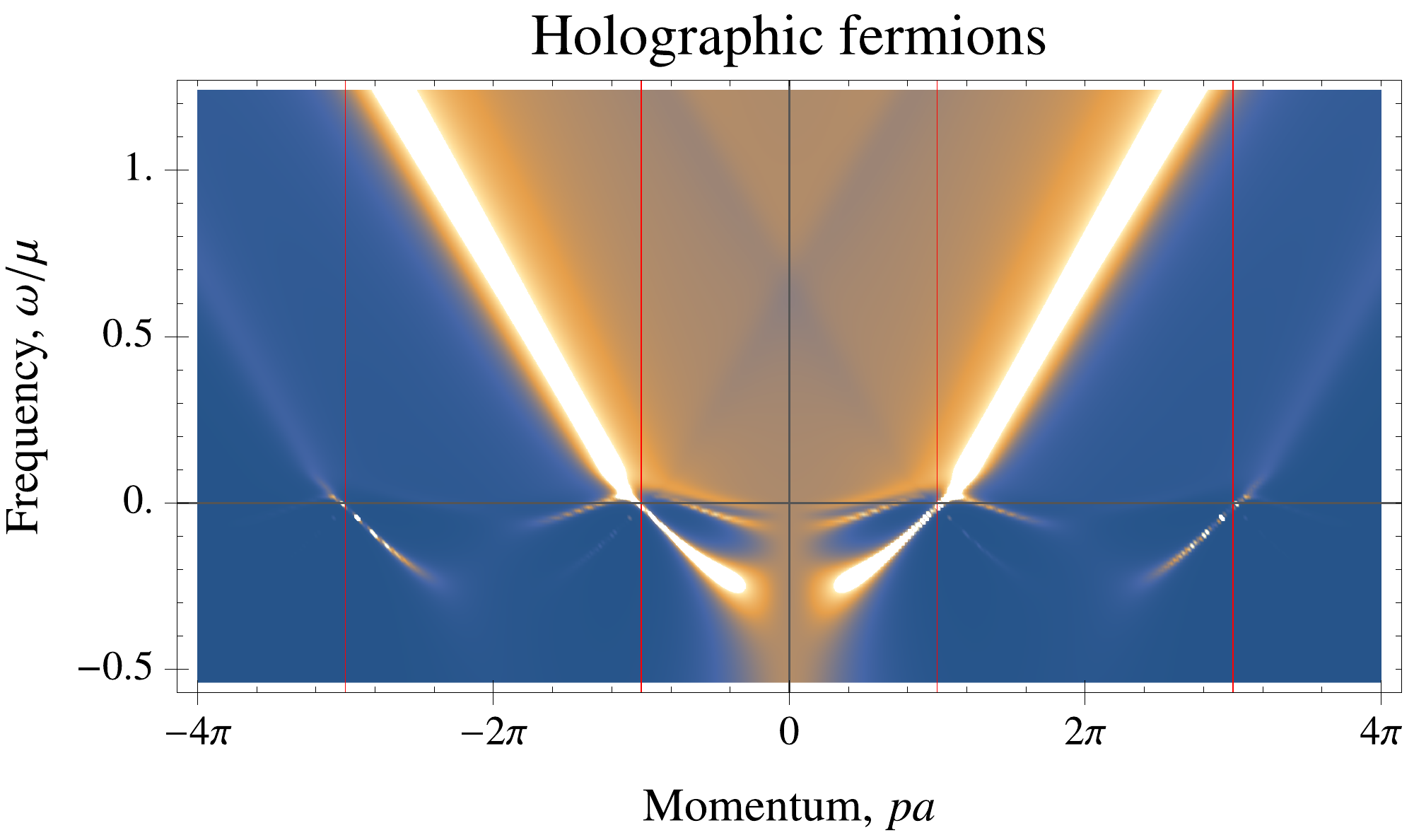}
\caption{\label{fig:ChainFig}  \label{fig:HolographicFig} \textbf{Lines of isolated zeros}. Spectral density of a fermion is shown in color. The Brillouin zone boundaries are shown with red grid lines. The bright bands correspond to the dispersion relation of itinerant fermion $\xi_p$, while the finite background is due to the critical continuum. The lines of zeros are seen as darker bands and correspond to the dispersion of the Umklapp copies of the fermionic dispersion $\xi_{p \pm 2 \pi/a}$.
\textbf{\textit{Left panel:} In SYK chain.} We use free space dispersion $\xi_p = p^2/2m-\mu$ with $\mu=0.2, m = 0.5$, SYK Green's function \eqref{GR_SYK} has $J=10$, and the coupling is $\lambda=0.3$. The gray dashed lines show the dispersions $\xi_{p \pm \Delta p}$.
\textbf{\textit{Right panel:} In the holographic model for fermions on top of the continuum.} The model has periodic background with period $a$.
}
\end{figure*}

This isolated zero phenomenon in the case of a finite number of spatial dimensions can be illustrated in a holographic model of a strange metal. \cite{zaanen2015holographic} In this framework one describes strongly entangled quantum systems in terms of a classical gravitational problem with one extra dimension, see Fig. \ref{fig:holo}. At finite chemical potential these systems can flow to a novel ``strange metal'' low energy sector that exhibits local quantum criticality, i.e. it contains a continuum sector. The fermionic spectral function can be computed from the propagator of a dual holographic fermion $\mathcal{X}$ in the curved space with a charged black hole horizon. We refer the reader to the seminal works. \cite{vcubrovic2009string,faulkner2010strange,faulkner2011emergent} Importantly, the dynamics of a holographic fermion can be understood as a momentum dependent probe of the quantum critical continuum sector. In holographic models this sector is local, i.e. the system falls apart into local domains, each of which is a continuum theory in its own right. \cite{iqbal2012semi} The SYK model is an explicit proposal for the microscopic theory of one such domain.  
This explains why at low energy this holographic theory is similar to~\eqref{H_chain} with the SYK Hamiltonian replaced by the more abstract local quantum critical Hamiltonian. \cite{faulkner2011semi} The Green's function of the critical fermions $\psi$ can be shown to have the scaling form $\mathcal{G}(\omega) \sim \omega^{2 \nu_p}$, where $\nu_p$ depends on the momentum of the itinerant probe $\chi_p$.

Unlike the case with an SYK chain, the holographic continuum comes from translationally invariant horizon, therefore it does not introduce the Brillouin zone by itself.
The weak interaction between the localized continuum sectors effectively restores translational invariance. 
The appearance of isolated zeros 
in the holographic spectrum becomes apparent once one breaks translations 
by introducing a crystal lattice with period $a$. 
The technical details of this construction are discussed in the specialized papers. \cite{liu2012lattice,ling2013holographic, cremonini2018holographic,Krikun_Balm} Importantly, in this case it is the periodic potential, which introduces the Bloch momentum and makes the dispersion $\xi_p$ multivalued due to the appearance of the $\xi_{p- 2\pi/a}$ and $\xi_{p+ 2\pi/a}$ copies. These branches are coupled by the near horizon continuum $\mathcal{G}(\omega)$, and this leads to the similar pattern in the spectral function as in the SYK chain case, see Fig. \ref{fig:HolographicFig}, right panel. At finite temperature the zeros are smeared and the spectral density never strictly vanishes. However the distinctive line shape is clearly recognizable.

\section{Conclusion}
We have observed a characteristic phenomenon which arises in quantum systems having both discrete spectrum and continuum subparts. The interference between the discrete parts mediated by 
the continuum gives rise to the isolated zeros in the spectral function of the one discrete state, which are located at the energy levels of the other states. It is the complement of the known Fano resonance. We derive this effect in full generality and show that it is present for a very general class of the continuum subsystems, provided their Green's functions have finite imaginary part and lack a pole structure. 

By studying several examples with and without spatial dimensions, we see that the phenomenon is generally present. 
It is characterized by the distinctive ``pole-zero'' line shape, which is different from that of the Fano resonance, since the positions of the pole and zero are in principle independent of each other. We show that this line shape is not destroyed by the additional couplings or disorder: The exact zero gets smeared, but the localized depression remains.
This leads us to propose that our finding can serve as a convenient experimental marker for a quantum system to include a continuum subpart, such as an SYK continuum in the graphene flake on a quantum dot or a local quantum critical phase in case of a strange metal.

\section*{Acknowledgment} We benefited from discussions with 
\.{I}nan\c{c} Adagideli, Sa\v{s}o Grozdanov, Aurelio Romero-Berm{\'u}dez, Floris Balm, Jimmy Hutasoit, and Andrei Pavlov.
This research was supported in part by a VICI award of the Netherlands Organization for Scientific Research (NWO), by the Netherlands Organization for Scientific Research/Ministry of Science and Education (NWO/OCW), by the Foundation for Research into Fundamental Matter (FOM),  and by an ERC Synergy Grant. A.K. and K.S. also wish to thank the organizers and participants of the workshop {\em Many-body Quantum Chaos, Bad Metals and Holography} made possible by support from NORDITA, ICAM (the Institute for Complex Adaptive Matter) and Vetenskapradet.

\newpage

\bibliography{zeros}

\begin{widetext}
\appendix

\section{Inclusion of general coupling between discrete levels}
\label{app_kappa}

As in the main text, we consider two fermions coupled through continuum but with an additional direct coupling: 
\begin{align}
S&{}=S_\chi+S_\psi+S_{int} \label{appA_S}, \\ \nonumber
S_\chi&{}=\int_0^\beta d \tau
\begin{pmatrix}
\bar{\chi}_{A} & \bar{\chi}_{B}
\end{pmatrix}
\begin{pmatrix}
\partial_\tau +\Omega_A  & \kappa \\  \kappa^* &  \partial_\tau +\Omega_B
\end{pmatrix}
\begin{pmatrix}
\chi_{A} \\ \chi_{B}
\end{pmatrix}  , \\  \nonumber
S_\psi&{}=-\int_0^\beta d \tau \int_0^\beta d \tau' \, \bar{\psi}(\tau) \mathcal{G}\left(\tau-\tau'\right)^{-1} \psi(\tau') , \\  \nonumber
S_{int}&{}=\int_0^\beta d \tau \sum_{\sigma=A,B} \left(\lambda_\sigma \,  \bar{\psi} \, \chi_{\sigma}+\lambda_\sigma^* \bar{\chi}_{\sigma} \psi\right),
\end{align}
where $\kappa$ is the direct coupling strength.
After integration over the fermion $\psi$, we derive the inverse Green's function for the fermions $\chi_A$, $\chi_B$:
\begin{align} 
G_{\sigma\sigma'}(\mathrm{i}\omega_n)^{-1}&{}= \begin{pmatrix}
\mathrm{i} \omega_n -\Omega_A -|\lambda_A|^2  \, \mathcal{G}(\mathrm{i}\omega_n) & -\kappa -\lambda_A^*\lambda_B  \, \mathcal{G}(\mathrm{i}\omega_n) \\
-\kappa^* -\lambda_A \lambda_B^* \, \mathcal{G}(\mathrm{i}\omega_n) & \mathrm{i} \omega_n -\Omega_B -|\lambda_B|^2  \, \mathcal{G}(\mathrm{i}\omega_n) 
\end{pmatrix} \label{appA_Gm1}, 
\end{align}
where $\omega_n=\pi T\left(2 n+1\right)$ are Matsubara frequencies. Inversion of the matrix (\ref{appA_Gm1}) and analytic continuation $\mathrm{i} \omega_n \to \omega +\mathrm{i} \delta$ with $\delta=0^+$ gives the retarded Green's function: 
\begin{align} 
G^R_{\sigma\sigma'}(\omega)&{}\!=\!\frac{1}{D(\omega)}  \begin{pmatrix}
\omega -\Omega_B +\mathrm{i} \delta -|\lambda_B|^2  \, \mathcal{G}^R(\omega) & \kappa +\lambda_A^*\lambda_B  \, \mathcal{G}^R(\omega) \\
\kappa^* +\lambda_A \lambda_B^* \, \mathcal{G}^R(\omega) & \omega -\Omega_A +\mathrm{i} \delta -|\lambda_A|^2  \, \mathcal{G}^R(\omega) 
\end{pmatrix} \label{appA_GR}, \\
D(\omega)&{}\!=\!\left(\omega \!-\!\Omega_A \!+\!\mathrm{i} \delta\right)\left(\omega \!-\!\Omega_B\! +\!\mathrm{i} \delta\right)\!-\!|\kappa|^2\!-\!\left(|\lambda_A|^2\left(\omega\!-\!\Omega_B\! +\!\mathrm{i} \delta\right)\!+|\lambda_B|^2\left(\omega\!-\!\Omega_A \!+\!\mathrm{i} \delta\right)\!+\!\kappa \lambda_A \lambda_B^*\!+\!\kappa^* \lambda_A^* \lambda_B\right) \mathcal{G}^R(\omega) \label{appA_D}.
\end{align}
The spectral function for fermion $\chi_A$ is defined as the imaginary part of the $AA$ block of (\ref{appA_GR}):
\begin{align} \nonumber
A_A(\omega)&{}=-\frac{1}{\pi} \mathrm{Im} G^R_{AA}(\omega)=\\&{}= -\frac{|\lambda_A|^2}{\pi} \frac{  |\omega-\Omega_B +\kappa \, \frac{\lambda_A \lambda_B^*}{|\lambda_A|^2}|^2 \, \mathrm{Im} \mathcal{G}^R(\omega)}{\left|\left(\omega \!-\!\Omega_A\right)\left(\omega \!-\!\Omega_B\right)\!-\!|\kappa|^2\!-\!\left(|\lambda_A|^2\left(\omega\!-\!\Omega_B\right)\!+|\lambda_B|^2\left(\omega\!-\!\Omega_A\right)\!+\!\kappa \lambda_A \lambda_B^*\!+\!\kappa^* \lambda_A^* \lambda_B\right) \mathcal{G}^R(\omega)\right|^2}  \label{appA_DOSab}.
\end{align}
The imaginary part of the continuum Green's function $\mathcal{G}^R$ is supposed to be finite: $\mathrm{Im} \mathcal{G}^R \gg \delta=0^+$, so that $\delta$ can be neglected in (\ref{appA_DOSab}). 
Zeros of the spectral function (\ref{appA_DOSab}) are given by  solutions of the equation $(\omega-\Omega_B)^2+2\mathrm{Re}\zeta \, (\omega-\Omega_B)+|\zeta|^2=0$, where $\zeta = \kappa \lambda_A \lambda_B^*/|\lambda_A|^2$. In case of the complex valued $\zeta=|\zeta|\mathrm{e}^{\mathrm{i}\varphi}$, the solutions are $\omega =\Omega_B-|\zeta|\cos \varphi \pm |\zeta|\sqrt{\cos^2 \varphi - 1}$. This results in the appearance of shifted isolated zeros at the real energy axis only for $\varphi=\pi m$: $\omega=\Omega_B+\left(-1\right)^{m+1} |\zeta|$, where $m$ are integers. So, zeros of the spectral function (\ref{appA_DOSab}) are stable for real values of $\zeta$. 

In case of negligibility of the direct coupling $\kappa$ between the isolated states on the background of the frequencies of those, we restore the expression for the spectral function from the main text:
\begin{align} 
A_A(\omega)&{}= -\frac{|\lambda_A|^2}{\pi} \frac{  \left(\omega-\Omega_B\right)^2 \, \mathrm{Im} \mathcal{G}^R(\omega)}{\left|\left(\omega \!-\!\Omega_A\right)\left(\omega \!-\!\Omega_B\right)\!-\!\left(|\lambda_A|^2\left(\omega\!-\!\Omega_B\right)\!+|\lambda_B|^2\left(\omega\!-\!\Omega_A\right)\right) \mathcal{G}^R(\omega)\right|^2}  \label{appA_DOSa}.
\end{align}
The result (\ref{appA_DOSa}) is valid for general couplings $\lambda_\sigma$.


\section{Mean-field treatment of the SYK model with two component impurity}\label{app_SYK4}

The Euclidean action for the Hamiltonian (8) in the main text after disorder averaging is
\begin{align} \nonumber
S=&{}\int_0^\beta d\tau \bigg[ \sum_{\sigma=A,B} \bar{\chi}_\sigma\left(
\partial_\tau+\Omega_\sigma\right)  \chi_\sigma + \sum_{i=1}^N \bar{\psi}_i \, \partial_\tau \psi_i \bigg]\\&{}-\int_0^\beta d\tau \int_0^\beta d\tau' \bigg[\,\frac{\lambda^2}{N} \sum_{i=1}^N \sum_{\sigma,\sigma'=A,B} \bar{\chi}_\sigma \psi_i(\tau)\, \bar{\psi}_i \chi_{\sigma'}(\tau') +\frac{J^2}{4 N^3} \sum_{i,j,k,l=1}^N \bar{\psi}_i \bar{\psi}_j \psi_k \psi_l(\tau) \bar{\psi}_l \, \bar{\psi}_k \psi_j \psi_i(\tau')\bigg] \label{appB_Sdisav}.
\end{align} 
Introduction of two pairs of nonlocal fields:
\begin{align} \nonumber
1_\psi&{}=\int \mathcal{D} G_\psi \, \delta\left(G_\psi(\tau',\tau)-\frac{1}{N}\sum_{i=1}^N \bar{\psi}_i(\tau) \psi_i(\tau')  \right)= \\ &{}=\int \mathcal{D} \Sigma_\psi \int \mathcal{D} G_\psi \exp\Bigg[ N \int_0^\beta d \tau \int_0^\beta d \tau' \, \Sigma_\psi(\tau,\tau') \left(G_\psi(\tau',\tau)-\frac{1}{N}\sum_{i=1}^N \bar{\psi}_i(\tau) \psi_i(\tau')\right)\Bigg] \label{appB_1_psi}
\end{align}  
and
\begin{align} \nonumber
1_\chi&{}=\int \mathcal{D} G_\psi \, \delta\left(G_\chi(\tau',\tau)-\sum_{\sigma,\sigma'=A,B} \bar{\chi}_\sigma(\tau) \chi_{\sigma'}(\tau')  \right)= \\&{}=\int \mathcal{D} \Sigma_\chi \int \mathcal{D} G_\chi \exp\Bigg[\int_0^\beta d \tau \int_0^\beta d \tau' \, \Sigma_\chi(\tau,\tau')  \left(G_\chi(\tau',\tau)-\sum_{\sigma,\sigma'=A,B}\bar{\chi}_\sigma(\tau) \chi_{\sigma'}(\tau')\right)\Bigg] \label{appB_1_phi}
\end{align}   
allows us to rewrite the action (\ref{appB_Sdisav}) as
\begin{align} \nonumber
S=&{}\int_0^\beta d\tau \int_0^\beta d\tau' \Bigg[\sum_{\sigma,\sigma'=A,B} \bar{\chi}_\sigma(\tau)\left(\delta_{\sigma\sigma'}\delta(\tau-\tau')\left(
\partial_\tau+\Omega_\sigma\right) +\Sigma_\chi(\tau,\tau')  \right) \chi_{\sigma'}(\tau') +
\sum_{i=1}^N \bar{\psi}_i(\tau)\left(\delta(\tau-\tau')
\partial_\tau+\Sigma_\psi(\tau,\tau')  \right) \psi_i(\tau') 
\\&{}-\left(\Sigma_\chi(\tau,\tau')-\lambda^2 G_\psi(\tau,\tau')\right)G_\chi(\tau',\tau)-N \left(\Sigma_\psi(\tau,\tau')G_\psi(\tau',\tau)+\frac{J^2}{4}G_\psi(\tau,\tau')^2 G_\psi(\tau',\tau)^2 \right)\Bigg] \label{appB_SGS}.
\end{align} 
Assuming that all nonlocal fields $G_\psi$, $\Sigma_\psi$, $G_\chi$, and $\Sigma_\chi$ are the functions of $\tau-\tau'$, we get variational saddle-point equations:
\begin{align}
G_\psi(\tau-\tau')&{}= -\frac{1}{N} \sum_{i=1}^N \left\langle \mathrm{T}_\tau \psi_i(\tau)\bar{\psi}_i(\tau') \right\rangle \label{appB_eq_psi_1},\\
\Sigma_\psi(\tau-\tau')&{}=-J^2 G_\psi(\tau-\tau')^2 G_\psi(\tau'-\tau)+ \frac{\lambda^2}{N} G_\chi(\tau-\tau')\label{appB_eq_psi_2},\\
G_\chi(\tau-\tau')&{}= -\sum_{\sigma,\sigma'=A,B} \left\langle \mathrm{T}_\tau \chi_\sigma(\tau)\bar{\chi}_{\sigma'}(\tau') \right\rangle \label{eq_phi_1},\\
\Sigma_\chi(\tau-\tau')&{}=\lambda^2 G_\psi(\tau-\tau') \label{appB_eq_phi_2},
\end{align}
where  $G_\chi=\sum_{\sigma\sigma'}G_{\sigma\sigma'}$.
In the large $N$  and long time limit $1 \ll J \tau \ll N$, equations (\ref{appB_eq_psi_1}) and (\ref{appB_eq_psi_2}) are simplified to $G_\psi(\mathrm{i}\omega_n)^{-1}=-\Sigma_\psi(\mathrm{i}\omega_n)$ and $\Sigma_\psi(\tau-\tau')=-J^2 G_\psi(\tau-\tau')^2 G_\psi(\tau'-\tau)$ with a known zero temperature solution: \cite{sachdev2015bekenstein}
\begin{align}
G_\psi(\mathrm{i}\omega_n)=-\mathrm{i} \pi^{1/4} \frac{\mathrm{sgn}(\omega_n)}{\sqrt{J |\omega_n|}} \label{appB_G_SYK}.
\end{align}
Thus, we derived an effective action for the impurity fermions:
\begin{align}
S=\sum_{n=-\infty}^{+\infty}
\begin{pmatrix}
\bar{\chi}_{n,A} & \bar{\chi}_{n,B}
\end{pmatrix}
\begin{pmatrix}
-\mathrm{i} \omega_n +\Omega_A + \lambda^2 G_\psi(\mathrm{i}\omega_n) & \lambda^2 G_\psi(\mathrm{i}\omega_n) \\   \lambda^2 G_\psi(\mathrm{i}\omega_n) &  -\mathrm{i} \omega_n +\Omega_B +\lambda^2 G_\psi(\mathrm{i}\omega_n)
\end{pmatrix}
\begin{pmatrix}
\chi_{n,A} \\ \chi_{n,B}
\end{pmatrix}\, \label{appB_S_eff},
\end{align}
so that their Green's function is
\begin{align}
G_{\sigma\sigma'}(\mathrm{i}\omega_n)&{}=\frac{1}{D(\mathrm{i}\omega_n)} \begin{pmatrix}
\mathrm{i} \omega_n -\Omega_B -\lambda^2 G_\psi(\mathrm{i}\omega_n) & \lambda^2 G_\psi(\mathrm{i}\omega_n)\\
\lambda^2 G_\psi(\mathrm{i}\omega_n) & \mathrm{i} \omega_n -\Omega_A -\lambda^2 G_\psi(\mathrm{i}\omega_n)
\end{pmatrix} \label{appB_G},\\
D(\mathrm{i}\omega_n)&{}=\left(\mathrm{i}\omega_n -\Omega_A\right)\left(\mathrm{i}\omega_n -\Omega_B\right)-\lambda^2\left(2\mathrm{i} \omega_n -\Omega_A -\Omega_B\right)G_\psi(\mathrm{i}\omega_n)\, \label{appB_Dm}.
\end{align}
We perform analytic continuation $\mathrm{i}\omega_n \to \omega+\mathrm{i} \delta$ with $\delta=0^+$ to derive the retarded Green's function:
\begin{align} 
G^R_{\sigma\sigma'}(\omega)=&{}\frac{1}{D(\omega)} \begin{pmatrix}
\omega -\Omega_B +\mathrm{i} \delta-\lambda^2 G^R_\psi(\omega) & \lambda^2 G^R_\psi(\omega)\\
\lambda^2 G^R_\psi(\omega) &\omega -\Omega_A+\mathrm{i} \delta -\lambda^2 G^R_\psi(\omega)
\end{pmatrix} \label{appB_GR}, \\
D(\omega)&{}= \left(\omega -\Omega_A+\mathrm{i} \delta\right)\left(\omega -\Omega_B+\mathrm{i} \delta\right)-\lambda^2\left(2\omega -\Omega_A -\Omega_B+2\mathrm{i} \delta\right)G^R_\psi(\omega) \label{appB_D},
\end{align}
where 
\begin{align}
G^R_\psi(\omega)=-\mathrm{i} \pi^{1/4}\frac{\mathrm{e}^{\mathrm{i} \pi \mathrm{sgn}(\omega)/4}}{\sqrt{J|\omega|}} \label{appB_GR_SYK},
\end{align}
that fulfills $\mathrm{Im}G^R_\psi \gg \delta=0^+$. 
The spectral function of the fermion $\chi_A$ is given by the imaginary part of the corresponding matrix element of (\ref{appB_GR}) 
\begin{align}
A_A(\omega)=-\frac{1}{\pi} \mathrm{Im} G^R_{AA}(\omega)=-\frac{\lambda^2}{\pi} \frac{\left(\omega-\Omega_B\right)^2 \mathrm{Im} G^R_\psi(\omega)}{\left|\left(\omega -\Omega_A\right)\left(\omega -\Omega_B\right)-\lambda^2\left(2\omega -\Omega_A -\Omega_B\right)G^R_\psi(\omega)\right|^2}  \label{appB_DOSab},
\end{align}
where $\delta=0^+$ is neglected.
The result (\ref{appA_DOSab}) coincides with the expression (3) in the main text for $\mathcal{G}^R\equiv G^R_\psi$.

\section{Multiple states coupled to the SYK continuum}\label{app_ManyStates}

Here we address the case of $M$ discrete states $\sigma=A, B_1, \ldots, B_{M-1}$ coupled to the SYK continuum. To stay in the non-Fermi liquid regime we suppose $M\ll N$. \cite{banerjee2017solvable} Once SYK degrees of freedom are integrated out, the effective action for $\chi_\sigma$ fermions is similar to (\ref{appB_S_eff}) in Appendix \ref{app_SYK4}:
\begin{align} \nonumber
S=&{}\sum_{n=-\infty}^{+\infty} \sum_{\sigma,\sigma'=A,B_1,\ldots,B_{M-1}} \bar{\chi}_{n,\sigma} \left( \delta_{\sigma\sigma'}\left(-\mathrm{i}\omega_n + \Omega_\sigma\right) + \lambda^2 G_\psi(\mathrm{i} \omega_n) \right) \chi_{n,\sigma'}=\\ \nonumber =&{}\sum_{n=-\infty}^{+\infty} \bigg[\sum_{\sigma,\sigma'=A,B_1,\ldots,B_{M-2}} \bar{\chi}_{n,\sigma} \left(\delta_{\sigma\sigma'}\left(-\mathrm{i}\omega_n + \Omega_\sigma\right) + \lambda^2 G_\psi(\mathrm{i} \omega_n) \right) \chi_{n,\sigma'}+ \\ \nonumber &{}+ \bar{\chi}_{n,B_{M-1}} \left( -\mathrm{i}\omega_n + \Omega_{B_{M-1}} + \lambda^2 G_\psi(\mathrm{i} \omega_n) \right)  \chi_{n,B_{M-1}} + \sum_{\sigma=A,B_1,\ldots,B_{M-2}} \lambda^2 G_\psi(\mathrm{i} \omega_n) \left(\bar{\chi}_{n,B_{M-1}}\chi_{n, \sigma}+\bar{\chi}_{n, \sigma}\chi_{n,B_{M-1}}\right)\bigg]=\\\nonumber = &{}\sum_{n=-\infty}^{+\infty} \sum_{\sigma,\sigma'=A,B_1,\ldots,B_{M-2}} \bar{\chi}_{n,\sigma} \left(\delta_{\sigma\sigma'}\left(-\mathrm{i}\omega_n + \Omega_\sigma\right) + \lambda^2  \underbrace{\frac{\left(-\mathrm{i} \omega_n + \Omega_{B_{M-1}}\right)G_\psi(\mathrm{i} \omega_n)}{-\mathrm{i} \omega_n + \Omega_{B_{M-1}} + \lambda^2 G_\psi(\mathrm{i} \omega_n)}}_{\equiv G_{M-1}(\mathrm{i} \omega_n)}\right) \chi_{n,\sigma'} = \\ \nonumber =&{} \sum_{n=-\infty}^{+\infty} \sum_{\sigma,\sigma'=A,B_1,\ldots,B_{M-3}} \bar{\chi}_{n,\sigma} \left(\delta_{\sigma\sigma'}\left(-\mathrm{i}\omega_n + \Omega_\sigma\right) + \lambda^2  \underbrace{\frac{\left(-\mathrm{i} \omega_n + \Omega_{B_{M-2}}\right)G_{M-1}(\mathrm{i} \omega_n)}{-\mathrm{i} \omega_n + \Omega_{B_{M-2}} + \lambda^2 G_{M-1}(\mathrm{i} \omega_n)}}_{\equiv G_{M-2}(\mathrm{i} \omega_n)}\right) \chi_{n,\sigma'} =  \ldots = \\  =&{} \sum_{n=-\infty}^{+\infty}  \bar{\chi}_{n,A} \left(-\mathrm{i}\omega_n + \Omega_A + \lambda^2  \underbrace{\frac{\left(-\mathrm{i} \omega_n + \Omega_{B_1}\right) G_2(\mathrm{i} \omega_n)}{-\mathrm{i} \omega_n + \Omega_{B_1} + \lambda^2 G_2(\mathrm{i} \omega_n)}}_{\equiv G_1(\mathrm{i}\omega_n)} \right) \chi_{n,A} \, \label{appC_S},
\end{align}
where the derivation of the last line in (\ref{appC_S}) is based on the one-by-one Gaussian integration over all states except $\chi_A$ and $G_\psi$ is given by the expression (\ref{appB_G_SYK}).
Now we restore $G_1$ back:
\begin{align}\nonumber
G_1(\mathrm{i}\omega_n)=&{}\frac{\left(-\mathrm{i} \omega_n + \Omega_{B_1}\right) G_2(\mathrm{i} \omega_n)}{-\mathrm{i} \omega_n + \Omega_{B_1} + \lambda^2 G_2(\mathrm{i} \omega_n)}=\frac{\left(-\mathrm{i} \omega_n + \Omega_{B_1}\right)\left(-\mathrm{i} \omega_n + \Omega_{B_2}\right) G_3(\mathrm{i} \omega_n)}{\left(-\mathrm{i} \omega_n + \Omega_{B_1}\right)\left(-\mathrm{i} \omega_n + \Omega_{B_2}\right) + \lambda^2 G_3(\mathrm{i} \omega_n) \left(-\mathrm{i} \omega_n + \Omega_{B_1}-\mathrm{i} \omega_n + \Omega_{B_2}\right)} = \\ \nonumber =&{} \frac{\prod_{m=1}^3\left(-\mathrm{i} \omega_n + \Omega_{B_m}\right) G_4(\mathrm{i} \omega_n)}{\prod_{m=1}^3\left(-\mathrm{i} \omega_n + \Omega_{B_m}\right) + \lambda^2 G_4(\mathrm{i} \omega_n) \sum_{m=1}^3 \prod_{p\neq m}^3 \left(-\mathrm{i} \omega_n + \Omega_{B_p}\right)}= \ldots =\\=&{} \frac{\prod_{m=1}^{M-1}\left(-\mathrm{i} \omega_n + \Omega_{B_m}\right) G_\psi(\mathrm{i} \omega_n)}{\prod_{m=1}^{M-1}\left(-\mathrm{i} \omega_n + \Omega_{B_m}\right) + \lambda^2 G_\psi(\mathrm{i} \omega_n) \sum_{m=1}^{M-1} \prod_{p\neq m}^{M-1} \left(-\mathrm{i} \omega_n + \Omega_{B_p}\right)}\, \label{appC_G1}.
\end{align}
Finally, 
\begin{align}
G_{AA}(\mathrm{i} \omega_n) = \frac{\prod_{m=1}^{M-1}\left(\mathrm{i} \omega_n - \Omega_{B_m}\right) - \lambda^2 G_\psi(\mathrm{i} \omega_n) \sum_{m=1}^{M-1} \prod_{p\neq m}^{M-1} \left(\mathrm{i} \omega_n - \Omega_{B_p}\right)}{\left(\mathrm{i} \omega_n - \Omega_A\right)\prod_{m=1}^{M-1}\left(\mathrm{i} \omega_n - \Omega_{B_m}\right)-\lambda^2 G_\psi(\mathrm{i} \omega_n)\left( \left(\mathrm{i} \omega_n - \Omega_A\right)\sum_{m=1}^{M-1} \prod_{p\neq m}^{M-1} \left(\mathrm{i} \omega_n - \Omega_{B_p}\right)+\prod_{m=1}^{M-1}\left(\mathrm{i} \omega_n - \Omega_{B_m}\right) \right)}\, \label{appC_G}.
\end{align}
The spectral function we are interested is 
\begin{align} 
A_A(\omega)&{}=-\frac{1}{\pi} \mathrm{Im} G_{AA}\left(\mathrm{i} \omega_n\to \omega+\mathrm{i} \delta\right) =-\frac{\lambda^2}{\pi}\frac{\mathrm{Im}G^R_\psi(\omega)}{\left|D(\omega) \right|^2} \prod_{m=1}^{M-1}\left(\omega - \Omega_{B_m}\right)^2  \label{appC_DOSab},  \\
D(\omega)&{}=\left(\omega - \Omega_A\right)\prod_{m=1}^{M-1}\left(\omega - \Omega_{B_m}\right)-\lambda^2 G^R_\psi(\omega)\left( \left( \omega - \Omega_A\right)\sum_{m=1}^{M-1} \prod_{p\neq m}^{M-1} \left(\omega - \Omega_{B_p}\right)+\prod_{m=1}^{M-1}\left(\omega - \Omega_{B_m}\right) \right)\, \label{appC_D}.
\end{align}
Result (\ref{appC_DOSab}) coincides with the two fermion case (\ref{appB_DOSab}) at $M=2$.

\section{Effective momentum coupling by the SYK chain}\label{app:umklapps}

Consider the action corresponding to the Hamiltonian of the SYK chain  one with one-dimensional propagating fermions $\chi$, which interact locally with the SYK nodes. After integrating out the SYK fermions one can write down the effective interaction term in the action as
\begin{equation}
V_{\rm eff} = \int dx dx' \lambda^2 \bar{\chi}(x) \mathcal{G}(x,x'|\omega) \chi(x').
\end{equation}
Locality of the interaction immediately means that 
\begin{equation}
\mathcal{G}(x,x'|\omega) = \delta(x-x') \rho(x) \mathcal{G}(\omega),
\end{equation}
where we introduce the ``position density of nodes'' $\rho(x)$ which is a collection of periodically distributed delta functions in the case of a chain $\rho(x) = \sum_n \delta(x + n a)$.
Moving to momentum space gives
\begin{equation}
V_{\rm eff} = \int dp \, d\Delta p \, \lambda^2 \bar{\chi}_{p} \chi_{p+\Delta p} \, \rho(\Delta p) \mathcal{G}(\omega),
\end{equation}
where $\rho(\Delta p)$ is simply a Fourier transform of $\rho(x)$. For the periodically distributed chain it has the form $\rho(\Delta p) = \sum_{n} \delta(\Delta p + 2\pi n/a)$, which brings the interaction term to the form used in expression (\ref{Gpp}) of the main text.
\begin{equation}
V_{\rm eff} = \int dp \sum_{n} \lambda^2 \bar{\chi}_{p} \chi_{p-  2 \pi n/a} \mathcal{G}(\omega).
\end{equation}

\end{widetext}

\end{document}